\documentclass[12pt]{article}
\usepackage[top=1.2in,bottom=1.2in,left=1in,right=1in]{geometry}  % set the margins to 1in on all sides
\usepackage{graphicx}              % to include figures
\usepackage{amsmath}               % great math stuff
\usepackage{amsfonts}              % for blackboard bold, etc
\usepackage{amsthm}                % better theorem environments
\usepackage{caption,subcaption} 
\usepackage{multirow}
\usepackage{lineno,hyperref}
\usepackage{authblk}
\usepackage[justification=centering]{caption}

\newtheorem{defn}{Definition}
 
\title{\textbf{Modeling pages left blank in university examination: A resolution in higher education process}}

\author[a]{Suman K. Ghosh}
\author[b]{Subhradev Sen}

\affil[a,b]{\small{Alliance School of Business, Alliance University, Bengaluru, India.}}
\date{}
\begin{document}
\maketitle

\begin{abstract}
Trees are the main sources of paper production, in most of the cases, as far as the intellectual usages are concerned. However, our planet is lacking in that particular natural resource due to rapid growth of population, urbanization, and increased pollution, more importantly non-judicial utilization of such kind. Indian education sectors (schools, colleges, universities) utilize a major part in consumption of papers as a classical practice for conducting examinations and other documentation activities. Our attempt in this article is to investigate and provide an optimal estimate of the number of pages actually required in answer booklet in higher education sector. Truncated Poisson distribution is found to be the best fit for the data on number of pages left blank in an answer booklet after conduction of semester end examinations. To predict the outcome based on various factors such as, lines per pages, words per line, types of examinations etc. suitable regression modeling is performed. A real data set, collected over a period of one month, is been analyzed to illustrate the methods and conclusion is accomplished in the direction of cost reduction, saving of papers, and in turn, logical uses of natural resource to protect environmental interests.
\end{abstract}

\clearpage
\section{Introduction}
\label{sec:1}
It goes without saying that trees are the main sources of producing papers, until alternatives are proven to be exactly similar, that are utilized for many possible activities in day to day execution of us. Most of the paper mills are in existence for a long time and hence present technologies fall in a wide spectrum ranging from oldest to the most modern. In Indian scenario, the mills use a variety of raw material viz., wood, bamboo, recycled fibre, bagasse, wheat straw, rice husk, etc. In terms of share in total production, approximately $25\%$ are based on wood, $58\%$ on recycled fibre and $17\%$ on agro-residues. 
India's share in global paper demand is gradually growing as domestic demand is increasing at a steady pace while demand in the western nations is contracting. According to Indian Paper Mill Association, the domestic demand in India grew from $9.3$ million tonnes in financial year 2008-09 to $17$ million tonnes in financial year 2017-18 at a compound annual growth rate (CAGR) of $6.9\%$. The futuristic view is that growth in paper consumption would be in multiples of gross domestic product and hence an increase in consumption by one kg per capita would lead to an increase in demand of one million tonnes. Among five important demand drivers, a likely pick-up from the education sector is prominent one. Printing and writing segment demand is expected to grow at a CAGR of $4.2\%$ and reach $5.7$ million tonnes in financial year 2020-21 on the back of an anticipated pick-up from the education sector with improving literacy rates and growing enrollment as well as increasing number of schools and colleges.\\
Therefore, caring nature by reducing the usage of papers is obvious one can easily do, if not striving for a proper alternative that fulfill our need in every possible sense (see Skog and Nicholson, 1998; Manzardo et al. 2014). Although there are regular plantation of trees required to produce paper products (Rudel, 2009), there are several alternatives of non-judicial and unstructured ways of misutilization of the same.\\
The caring nature in paper usage is an indirect approach of caring by scientifically fulfilling our classical need of papers for examination systems. However, the following two facts are noted in connection to the improper paper utilization in examination systems at the different academic institutions in India. Firstly, students are gradually losing the capacity of writing in case of broad answer type questions, and secondly, the number of pages provided in the main answer scripts during examination are not scientifically matched with actual demand or requirement.\\
Our objective in the current investigation is three-fold:
\begin{enumerate}
	\item[(a)] To identify the distribution of unutilized papers in examination at higher education and to find an optimal setting for number of papers should be provided in an answer script. 
	\item[(b)] To find out possible effects due to other variables to the leaving papers blank in examination answer scripts who take the examination in a classical pattern.
	\item[(c)] To address the utilization maximization in view of cost constraints related to answer scripts used.
\end{enumerate}
For the purpose of fulfilling the above objectives, truncated Poisson distribution along with count regression procedure are applied for modeling supported by a real data illustration. A multiple linear discriminant analysis is also performed in view of grouping into important categories with the help of a real data.
The rest of the article is organized as follows. Truncated Poisson is described with its possible applications in section~\ref{sec:2}. Section~\ref{sec:3} deals with count regression models with emphasis given in right truncated Poisson model with mixed effects. In section~\ref{sec:4}, a real data set on the pages left blank at a semester end examination in a higher education institute in India is been analyzed as per the objectives of the research mentioned above and the corresponding results are discussed in dedicated subsections. Section~\ref{sec:5} discusses about maximization of a linear utility function of pages in answer scripts subject to certain cost constraints. Finally, the section~\ref{sec:6} concludes. 
\section{Truncated Poisson distribution}
\label{sec:2}
The Poisson distribution is a discrete probability distribution usually applied to the number of events occurring within a speciﬁed period of time or space. Theoretically, the possible values of a Poisson random variate is non-negative integers (including 0) and there is no upper limit a Poisson random variate can stop for. \\
The Poisson distribution is characterized by a single parameter, usually denoted by $\lambda$ $(>0)$ 
\begin{defn}
	A random variable $X$ is said to have a Poisson distribution with parameter $\lambda$ if its probability mass function (pmf) is of the form
	\begin{align}	
	 Pr[X=x]= \frac{\lambda^xe^{-\lambda}}{x!} \, \text{for}\,x=0,1,2,\ldots.
	 \end{align}
\end{defn} 
Numerous applications of Poisson distribution can be found in literature. Some well known applications could be, number of arrivals in a service queue during a specific time interval, number of accidents per month in a city, number of order received per week for a particular product, number of defects in a quality inspection, and number of printing mistakes per page in a book.
The wide applicability of Poisson distribution, however,  does not lower down its importance, rather newer applications and characterizations are found out in recent years, see Ahmed~(1991), Johnson et al.~(2005), for more details. Nevertheless, the Poisson distribution is successfully used for situations where some kind of counting is involved.
\\
Truncation in Poisson distribution arises when some specified values are not possible to record (in terms of process and not in terms of availability) either initially or at the end of a Poisson variate range. The former is known as left truncation, while the later is known as right truncation. The theoretical truncated Poisson distribution was introduced by Plackett~(1953). \\
Right truncation (omission of values exceeding a specified value $r$) can occur if the counting mechanism is unable to deal with large numbers or the counting process under consideration is bounded by a finite number.
\begin{defn}
	A random variable $X$ is said to have a right truncated Poisson distribution, right truncated at $r$ i.e. the realized values of $X$ is bounded at a specified positive integer $r$, with parameter $\lambda$ if its pmf is of the form 
	\begin{align}
	Pr[X=x]=\frac{\lambda^x}{x!}\left(\sum_{n=1}^{r}\frac{\lambda^j}{j!}\right)^{-1}, x=0,1,2,\ldots,r.
	\end{align}
	\end{defn} 
If $X_1,X_2,\ldots,X_n$ are $n$ independent and identically distributed random variables from right truncated Poisson, then the MLE $\hat{\lambda}$ of $\lambda$ satisfies the following equation:
\begin{equation}
\sum_{n=1}^{r} \frac{\left(\bar{x}-j\right)\hat{\lambda}^j}{j!}=0 
\end{equation} 
The simple estimator (Moore,~1954) is $\lambda^*=\sum\nolimits_{j}\frac{x_j}{m},$ \\
where $m$ is the number of values of $x$ that are less than $r-1$; this is an unbiased estimator of $\lambda$.

\section{The count regression models}
\label{sec:3}
Count data regression models are used for special cases in which the response variable takes count values. It represents the number of events that occur in a given time period.  Winkelmann~(1995) studied the number of live births over a speciﬁed age interval of the mother, where the interest was to analyze the variation in terms of the mother’s schooling, age, and household income. Another example of count modeling is studied by Cameron, Trivedi, Milne and Piggott~(1988), where they studied the number of times that individuals utilize a health service, such as visits to a doctor or days in the hospital in the past year. The most popular methods to model count data are Poisson and negative binomial regression (Saﬀari and Adnan, 2011). Poisson regression is the more popular of the two and is applied to various fields.
\subsection{Poisson regression models}
In many situations of practical interest the response variable in an experiment or observational study is a count that is assumed to follow the Poisson distribution. Therefore, a more suitable way to deal with count data is to use the Poisson distribution. The regression model that uses these kinds of option is called the Poisson regression or the Poisson log-linear regression model. For more details use of Poisson regression, one could refer to Frome~(1983), Lawless~(1987), Consul and Famoye~(1992), Lambert~(1992) and references therein. 
\subsection{Truncated Poisson regression models}
When the response variable follows a right truncated Poisson distribution, we use right truncated Poisson regression model. In our investigation, to model the number of pages left blank in the main answer booklet in semester end examinations, right truncated Poisson distribution is utilized owing to the fact that counting is restricted by the total number of available pages in main answer booklet. \\
There could be three different varieties for right truncated Poisson regression, namely, fixed effect model, random effect model, and mixed effect model.
We concentrate in right truncated Poisson regression model for fixed effect on the predictors and random effects for clusters of explanatory variables. Moreover, the random effects to follow a normal distribution with mean $0$ and variance $\sigma^2$. 
\subsubsection{Method of Estimation}
Suppose that we have a sample of $n$ observations $Y_1,Y_2,...,Y_n$ which can be treated as realizations of independent Poisson random variables, with $Y_i \sim Poi(\lambda_i)$ right truncated at $Y_i \leq r$, and suppose that we want to let the mean $\lambda_i$ depend on a vector of explanatory variables $\boldsymbol{x_i}$ and random effects. For the Poisson probability function, a model for count data truncated on the right at value $r$ can be expressed as 
\begin{align}
Pr(Y_i=y_i|Y_i \leq r)=\frac{Pr(Y_i=y_i)}{Pr(Y_i \leq r)}=\frac{\lambda_i^{y_i}}{\left(\sum_{k=0}^{r}\frac{\lambda_i^k}{k!}\right)y_i!}, i=1,2,\ldots,m,
\end{align}
where $m$ is the number of observation after truncation. \\ 
The standard assumption is to use the exponential mean parametrization, $$\lambda_i=exp(\boldsymbol{x_i}^T\beta + {z_i}^Tu_i), i=1,2,\ldots,n.$$
In this expression, $\boldsymbol{x_i}$ is a vector of covariates and $\boldsymbol{\beta}$ is a vector of parameters (fixed effect coefficients). The coefficient $\beta$ can be interpreted as average proportionate change in the conditional mean $E[Y_i|\boldsymbol{x_i}]$ for a unit change is $\boldsymbol{x_i}$. $\boldsymbol{Z}$ is a design matrix of random effects clusters and $\boldsymbol{u}$ is a vector of random effects for that. \\
In general matrix notation, we can write it as 
\begin{align}
\boldsymbol{\lambda}=exp(\boldsymbol{X\beta} + \boldsymbol{Zu}),
\end{align}
where\\ $\boldsymbol{X}$: Design matrix of order $n \times p$ for fixed effect explanatory variables\\
$\boldsymbol{\beta}$: Vector of fixed effect coefficients\\
$\boldsymbol{Z}$: Design matrix of order $n \times q$ for random effect explanatory variables (clusters/groups) \\
$\boldsymbol{u}$:  Vector of random effect coefficients \\
The method of hierarchical likelihood method of estimation (h-Likelihood) is used to obtain the values of regression coefficients. Let $Y_{ij} \left(i=1,...,m; j=1,...,n_i \right)$ be the observations of the response variable. Let $u_i$ be the unobserved random effect on the $i^{th}$ individual. We consider the model
\begin{equation}
Pr\left(Y_{ij}=y_{ij}|u_i,y_{ij}\leq r \right)=  \frac{\lambda_i^{y_i}}{\left(\sum_{k=0}^{r}\frac{\lambda_i^k}{k!}\right)y_i!} 
\end{equation}  
such that 
\begin{equation}
\lambda_{ij}=exp({x_{ij}}^T\beta + {z_{ij}}^Tu_i), i=1,2,\ldots,n.
\end{equation} 
We assume a normal distribution for the random effects 
\begin{equation}
u_i \sim Normal\left(0,\sigma^2 \right)
\end{equation}
Therefore, the h-likelihood (h) is defined by 
\begin{equation}
h = L_1 \left(\beta ; y|\boldsymbol{u} \right)+L_2 \left(\sigma^2,\boldsymbol{u} \right) 
\end{equation} 
where $ L_1 \left(\beta ; y|\boldsymbol{u} \right) $ is the logarithm of the conditional Poisson density function for the response $\boldsymbol{Y}$ given $\boldsymbol{u}$ with parameter $ \boldsymbol{\lambda}=exp(\boldsymbol{X\beta} + \boldsymbol{Zu}) $, and $L_2 \left(\sigma^2,\boldsymbol{u} \right)$ is the logarithm of the Normal density function for the random effect $\boldsymbol{u}$. Thus, 
\begin{flalign}
L_1 \left(\beta ; y|\boldsymbol{u} \right) 
&= \sum_{ij} \left[y_{ij}\ln{(\lambda_{ij})} - \log{(y_{ij}!)} - \ln{\sum_{k=0}^{r}\frac{\lambda_{ij}^k}{k!}} \right] \nonumber\\ 
& = \sum_{ij} \left[y_{ij}\left({x_{ij}}^T\beta + {z_{ij}}^Tu_i \right) - \ln{\sum_{k=0}^{r} \frac{\left( exp({x_{ij}}^T\beta + {z_{ij}}^Tu_i)\right)^{k}}{k!}} - \log{(y_{ij}!)} \right]
\end{flalign} 
and 
\begin{equation}
L_2 \left(\sigma^2,\boldsymbol{u} \right) = - \sum_{i} \left[\frac{\ln{(2\pi)}}{2}+\frac{\ln{(\sigma^2)}}{2}+\frac{{u_i}^2}{2\sigma^2} \right]
\end{equation}  
The maximum h-likelihood estimators (MHLEs) are obtained by solving the following equations,
\begin{flalign} 
\frac{\partial h}{\partial \beta_l} = \sum_{ij} \left[y_{ij} - \frac{1}{\sum_{k=0}^{r} \frac{\left( exp({x_{ij}}^T\beta + {z_{ij}}^Tu_i)\right)^{k}}{k!}} \sum_{k} \frac{\left( exp({x_{ij}}^T\beta + {z_{ij}}^Tu_i)\right)^{k}}{(k-1)!} \right]&x_{ijl} = 0\nonumber\\
&\text{for}\; l=1,\ldots,p 
\end{flalign}
and 
\begin{flalign} 
\frac{\partial h}{\partial u_i} = \sum_{j} \left[y_{ij} - \frac{1}{\sum_{k=0}^{r} \frac{\left( exp({x_{ij}}^T\beta + {z_{ij}}^Tu_i)\right)^{k}}{k!}} \sum_{k} \frac{\left( exp({x_{ij}}^T\beta + {z_{ij}}^Tu_i)\right)^{k}}{(k-1)!} \right]&z_{ij} - \frac{u_i}{\sigma^2} = 0\nonumber\\
&\text{for}\; i=1,\ldots,m.
\end{flalign}
Iterative techniques like, Fisher scoring or Newton-Raphson method can be used to obtain the estimators of the parameters. For more details on the method of estimation for truncated Poisson regression with normal random effects, one could refer to Suaiee~(2013).
\section{Application with real life data}
\label{sec:4}
\begin{figure}[h!]
		\centering
	\includegraphics[scale=0.6]{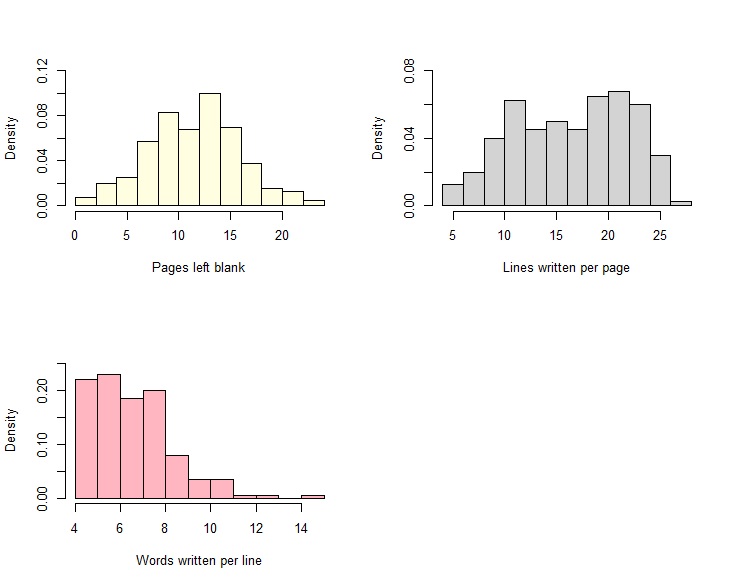}
	\caption{Histograms for response variable and predictors}
	\label{fig:histdata}
\end{figure}

\begin{figure}[h!]
\centering
\begin{subfigure}{.5\textwidth}
  \centering
  \includegraphics[width=.7\linewidth]{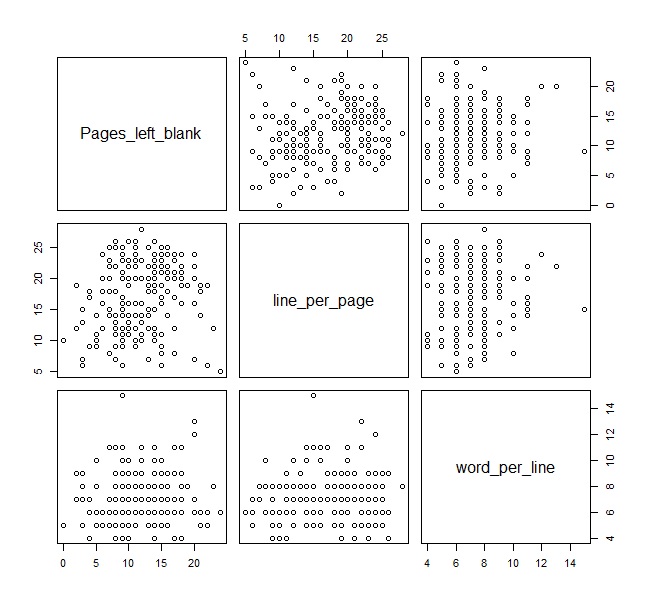}
  \caption{2D Scatter plot}
  \label{fig:2dplot}
\end{subfigure}%
\begin{subfigure}{.5\textwidth}
  \centering
  \includegraphics[width=.7\linewidth]{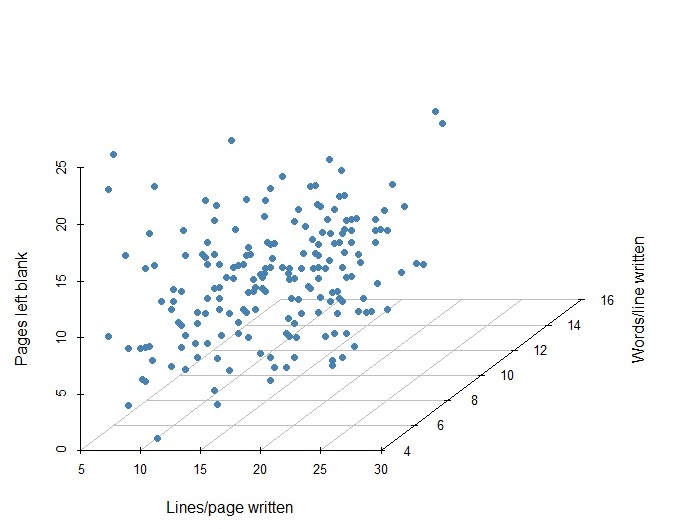}
  \caption{3D scatter plot}
  \label{fig:3dplot}
\end{subfigure}
\caption{Scatter plots for response variable and predictors}
\label{fig:scatterplot}
\end{figure}

This section illustrates the methods, described above, with the help of a real data analysis.
For the purpose, a sample of $200$ students appeared for semester end examination (SEE) are collected from a leading higher education institute in India during November-December, 2018. Students from various courses and subjects are been considered for balancing possible bias in sampling procedure. However, convenience sampling scheme were applied with adjustments in courses and paper types (quantitative and non-quantitative) for which SEE is taken by the students. Information on the following variables are collected:
\begin{enumerate}
	\item Course type (under graduate and post graduate).
	\item Type of paper written (quantitative and non-quantitative).
	\item Number of pages left blank\footnote{The total number of pages in main answer booklet is $25$ in the smaple collected, excluding front cover page-its immediate back page and one back cover page.}.
	\item Number of lines written per page\footnote{Number of lines per page is $29$ in the sample collected.}.
	\item Number of words written per line.
\end{enumerate}
For the last three variables, three random observations are taken to ensure unbiasedness and their average is considered.\\
Statistical software \textsf{R} (version 3.6.0) is utilized for calculations and we see that there are $24\%$ post graduate and $76\%$ undergraduate students in the sample. Quantitative paper was for $56\%$ and non-quantitative for $44\%$. From Fig.~\ref{fig:histdata}, we see that the variable pages left blank is normally distributed whereas words written per line is positively skewed. The scatter plots for response variable and predictors are displayed in Fig.~\ref{fig:scatterplot}. 

\subsection{Justification for using truncated Poisson}
\begin{figure}[h!]
		\centering
	\includegraphics[scale=0.5]{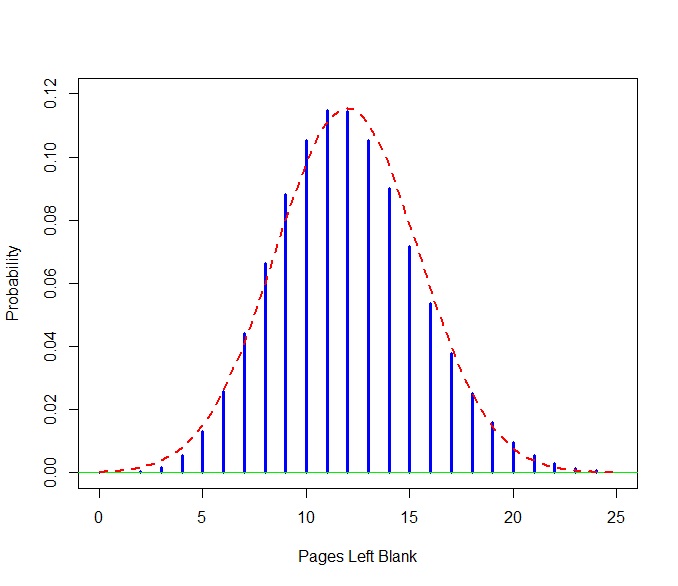}
	\caption{Data fitted with right truncated Poisson and normal curve}
	\label{fig:datafit}
\end{figure}

Before going to have certain model building on the response variable "pages left blank", let us have the justification for using truncated Poisson distribution (right truncated at 25, the maximum pages in an answer script). We fit the observations on the number of pages left blank with Poisson distribution (without truncation) and right truncated Poisson distributions, respectively. We use maximum likelihood (ML) method of estimation and fitted the Poisson and right Truncated Poisson distributions for the data on the variable ``number of pages left blank". As a model selection criteria, the following measures are considered.
\begin{enumerate}
	\item[(i)] Akaike information criteria (AIC): $AIC=2k-2\ln(\text{loglikelihood})$
	\item[(ii)] Consistent Akaike information criteria (cAIC): 	$cAIC=AIC+\frac{2k(k+1)}{n-k-1}$
	\item[(iii)] Bayesian information criteria (BIC): 	$BIC=k \ln(n)+2\ln(\text{loglikelihood})$
\end{enumerate}
Here, $n$: number of observations and $k$: number of parameters estimated. Lower the values of AIC, cAIC, and BIC, better is the fit. From Table~\ref{tab:MLest}, we observe that, right truncated (truncated at 25) Poisson distribution is better for the purpose of modeling.
We obtain (refer Table~\ref{tab:MLest}) expected number of pages left blank$=11.969\approx 12$.
From Fig.~\ref{fig:datafit}, we see that the pages left blank data is fitted with right truncated Poisson distribution.

\begin{table}[h]
	\centering
	\caption{ML estimates and model section measures}
	\label{tab:MLest}
	\scalebox{0.8}{
		\begin{tabular}{lccccc}
			\hline 
			Distribution &$\hat{\lambda}$(Std. Error) &AIC &cAIC &BIC \\
			\hline
			Poisson   &11.965(0.24459)
			&1215.955   &1215.975	&1219.253\\
			Right truncated Poisson  &11.969(0.24527) &1214.035   &1214.055	&1217.333\\   
			\hline 
		\end{tabular}
	}
\end{table}

\subsection{Truncated Poisson regression with different clusters}
In this section we consider response variable as ``number of pages left blank". The predictors or explanatory variables are taken as ``lines written per page" and ``number of words written per line" along with a general mean effect (intercept). We develop three right truncated Poisson regression models considering normal random effects for three different cluster types. 

\subsubsection{Model-A: Course types as clusters}
We consider course type classified as ``under-graduate" and  ``post-graduate" as different clusters having normal random effect. The predictors or explanatory variables are taken as ``lines written per page" and ``number of words written per line" along with a general mean effect (intercept). \\
Applying right truncated Poisson regression with normal random effects for course types as clusters, the result obtained is given in Table~\ref{tab2}. The log-likelihood, AIC, and BIC values for the model are obtained as $-786.5943$, $1579.189$, and $1589.084$, respectively. 
\begin{table}[h!]
	\centering
	\caption{Regression analysis table: Random effects for course type clusters}
	\label{tab2}
	\begin{tabular}{lcccc}
		\hline 
		Coefficients &Estimate (Std. Error) &t-value &P-value \\
		\hline
		Intercept   &2.03723(0.10307)  	&19.766  	&$<$ 0.0001\\ 
		Lines per page  &0.01666(0.00378)   &4.408 		&0.00010 \\   
		Words per line &0.00502(0.01106)   	&0.454     	&0.65000 \\    
		\hline 
	\end{tabular}
\end{table}

\subsubsection{Model-B: Each individual as cluster}
Next, we have considered each individual/student as different clusters having normal random effect. The predictors or explanatory variables are taken as ``lines written per page" and ``number of words written per line" along with a general mean effect (intercept).\\
Applying right truncated Poisson regression with normal random effects for individual clusters, the result obtained is given in Table~\ref{tab3}. The log-likelihood, AIC, and BIC values for the model are obtained as $-711.2469$, $1428.494$, and $1438.389$, respectively. 

\begin{table}[h]
	\centering
	\caption{Regression analysis table: Random effects for individual clusters}
	\label{tab3}
	\begin{tabular}{lcccc}
		\hline 
		Coefficients &Estimate (Std. Error) &t-value &P-value \\
		\hline
		Intercept   &2.10554(0.10326)  	&20.392  	&$<$ 0.0001\\ 
		Lines per page  &0.01349(0.00378)   &3.570 		&0.00036\\   
		Words per line &0.00650(0.01105)   	&0.588 		&0.55683\\    
		\hline 
	\end{tabular}
\end{table}
\subsubsection{Model-C: Types of paper written as clusters}
We next consider type of paper written (classified as quantitative and non-quantitative) as two different clusters having normal random effect. The predictors or explanatory variables are taken as ``lines written per page" and ``number of words written per line" along with a general mean effect (intercept). \\ 
Applying right truncated Poisson regression with normal random effects for clusters, the result obtained is given in Table~\ref{tab4}. The log-likelihood, AIC, and BIC values for the model are obtained as $-702.2191$, $1410.438$, and $1420.333$, respectively.
\begin{table}[h]
	\centering
	\caption{Regression analysis table: Random effects for paper type clusters}
	\label{tab4}
	\begin{tabular}{lcccc}
		\hline 
		Coefficients &Estimate (Std. Error) &t-value &P-value \\
		\hline
		Intercept   &2.13408(0.10324)  	&20.670  	&$<$0.0001\\ 
		Lines per page  &0.00768(0.00377)  	 &2.036  	&0.0418\\   
		Words per line &0.01012 (0.01104)   &0.916  	&0.3595\\    
		\hline 
	\end{tabular}
\end{table}
\begin{table}[h]
	\centering
	\caption{Model comparison and information measures}
	\label{tab5}
	\begin{tabular}{lcccc}
		\hline 
		Coefficients &-Log-likelihood &AIC &BIC \\
		\hline
		Model-A  &-786.5943   &1579.189  	&1589.084\\ 
		Model-B  &-711.2469   &1428.494  	&1438.389\\   
		Model-C  &-702.2191   &1410.438  	&1420.333\\    
		\hline 
	\end{tabular}
\end{table}
According to AIC and BIC values, Model-C (types of paper written as clusters) comes out as improved model (refer Table~\ref{tab5}). However, for each of the model words written per line is insignificant predictor.

\subsection{A linear discrimination approach of grouping}
In this section, our objective is to determine whether the variables i.e. pages left blank, lines written per page, and words written per line, will discriminate between quantitative and non-quantitative type paper. Discriminant analysis is a useful multivariate classification technique to predict membership in two or more mutually exclusive groups. We have used paper type (quantitative, non-quantitative) as grouping variable and pages left blank, lines per page, and words per line as independent variables. We have conducted Box's test of homogeneity of covariance matrices and obtained Box's M value as 13.592 which is significant with p-value, $p=0.038$, to conclude that the groups do differ in their covariance matrices. Wilks' lambda, a measure of how well the discriminant function separates cases into groups, is obtained as $0.543$ which is highly significant $(p < < 0.05)$. The small significance value indicates that the discriminant function does better than chance at separating the groups. The discriminant function is obtained as (considering standardized canonical discriminant function coefficients)
\begin{align}
D_i = 0.390 \times B_i + 0.923 \times L_i - 0.288 \times W_i,
\end{align}
where\\ 
$D_i$: Discriminant score for the $i^\text{th}$ student.\\
$B_i$: Number of pages left blank by the $i^\text{th}$ student. \\
$L_i$: Number of lines written per page by the $i^\text{th}$ student.\\
$W_i$: Number of words written per line by the $i^\text{th}$ student. \\

\begin{table}[h!]
	\centering
	\caption{Classification results}
	\label{tab:classification}
	\begin{tabular}{lcc}
		\hline 
		 &\multicolumn{2}{c}{Predicted membership}  \\
Actual membership $\downarrow$	&Quantitative   &Non-quantitative \\
\hline
Quantitative      &86        &26 \\
		                                 &$(72\%)$    &$(23\%)$ \\
Non-quantitative  &12        &76 \\ 
		                                 &$(14\%)$    &$(86\%)$ \\
\hline 
	\end{tabular}
\end{table}
The cut-off value of discriminant score is calculated by taking average of group centroids (mean discriminant score for each group) and is obtained as $18.79$. The model will classify any paper as quantitative if the discriminant score is less than $18.79$ and non-quantitative otherwise. 
For example, if we take a random observation i.e. an answer script having 9 pages left blank, 22 lines written per page and 7 words written per line; the discriminant score is obtained as $0.390 \times 9+0.923 \times 22-0.288 \times 7=21.8$, which means this answer script would be classified as a non-quantitative paper type. The classification result (i.e. actual versus predicted group membership) is shown in Table~\ref{tab:classification}, where the overall $81.5\%$ actual group cases are correctly classified.\\
The following important findings along with specific recommendations are noted in this section.
\begin{enumerate}
\item The expected number of pages left blank in main answer script is $12$, i.e., expected number of pages written is $13$. We recommend to utilize the residual pages that are not used in main answer scripts for producing additional answer sheets (each with 4 pages composition). The benefit in doing so is that there could be a reduction in making cost and wastage of pages would be minimized as additional sheets can be used whenever required.
\item Types of paper written came out as an important predictor for the response variable, pages left blank, and hence is a meaningful grouping in discrimination.
\item A cut off score of $18.79$ discriminates an answer script in two non-overlapping categories if certain minimal information is provided. 
\end{enumerate}
This next section discusses about a possible maximization of utility of pages in a single semester of any particular year. 
\section{Discussion on utility maximization}
\label{sec:5}
The main objective in this section is to discuss about a maximization aspect of the difference of page utility from current to the modified page numbers, subject to costs incurred for such modification and to identify the optimal reduction required in number of pages in answer script. 
An utility maximization problem can be framed as below:\\
We define,\\
$X$: Number of pages currently used in a main answer script.\\
$N_1$: Number of main answer script used in any examination.\\
$c_{11}$: Making cost per page for an answer script with $X$ number of pages.\\
$X_1$: Number of pages should be used (after reduction following the procedure described in section~\ref{sec:4}) in a main answer script.\\
$c_{12}$: Making cost per page for an answer script with $X_1$ number of pages.\\
$c_{21}$: Making cost per page for an additional answer script with $4$ number of pages.\\
$N_2$: Number of additional answer script used in any examination.\\
$c_{22}$: Per unit cost for making additional $(X-X_1)/4$ number of additional answer script.\\
\\
Assuming a linear function, let us now define the current and revised utility in terms of total pages that can be utilized in the whole process.\\
Current utility: $(N_1 X+N_2 4)$\\
Revised utility: $N_1 X_1+N_2 4+\frac{N_1(X-X_1)}{4}$\\
We want to 
\begin{align}
\label{utility}
\text{Maximize}\;U(X, X_1)=(N_1 X+N_2 4)-\left[N_1 X_1+N_2 4+\frac{N_1(X-X_1)}{4}\right]+k=\frac{3N_1(X-X_1)}{4}+k,
\end{align}
where $k$ is an integer constant and $N_1(X-X_1)\equiv k\,(\text{mod}\,4)$.\\
Subject to the constraints,
\begin{align}
&N_1 X c_{11}-N_1 X_1 c_{12} \geq A_0 \quad\text{(surplus cost inequation for main answer script)}\\
& 4 N_2 c_{21}+\frac{N_1(X-X_1)}{4} c_{22} \leq A_0 \quad \text{(cost inequation for additional answer script)}\\
&\text{with}\; X, X_1 \geq 0.
\end{align}
Here $A_0$ is amount of threshold benefit which is known or specified.\\
Now, for given values of $c_{ij}'s;i=1,2,j=1,2$ and known $N_1, N_2$, one can easily optimize (an integer programming problem) the function in (\ref{utility}) for $X$ and $X_1$. 
\section{Concluding remarks}
\label{sec:6}
This article provides a scientific way of allocating pages in main answer scripts in classical examination system in higher education sector. The study is restricted to one particular higher education institute in India. However, scope for investigations are open for multi-centric observations in the different educational institute in the same country and/or foreign institutes. Estimate for the number of pages blank will be an important investigation for multi-centric study as all the higher education institutes do not provide same number of pages in main answer scripts. We hope this article shall provide the authorities, all stake holders, and the student community, an alarming consciousness about the proper utilization of the pages used for education and thereby shall protect the environment thinking the large scale impact of the same to the environment. 

%% BIBLIOGRAPHY %%%%%
%\clearpage

\end{document}